\documentstyle[12pt]{article}
\begin{document}

\begin{center}
{\Large \bf Instant--Form Approach to Two--Body
Systems\footnote{Talk presented by V.E.T.\ at the Workshop on
Relativisic Approaches to Few-Body Systems, Groningen, July 21,
1997.} }\\[2mm]

{\large\em A.F.\ Krutov,$^a$ V.E.\
Troitsky\phantom{,}$^b$}\\[2mm]

{\small
$^a$Samara State University, 443011 Samara, Russia,\\
$^b$Nuclear Physics Institute, Moscow State University,
119899 Moscow, Russia}\\[2mm]

\end{center}
\begin{abstract}
We present a relativistic treatment of the problem of soft
electromagnetic structure by the modified instant form of
relativistic Hamiltonian dynamics. Our approach uses
relativistic parametrization and so picks out the relativistic
invariant quantities on each stage of the calculation.
The electromagnetic current matrix element satisfies the
current conservation law automatically.
We use relativistic modified impulse approximation.
It is constructed in relativistic invariant way.
For composite systems (including the spin 1 case) the
approach guarantees the uniqueness of the solution and it does
not use such concepts as "good" and "bad" current components.
The approach describes correctly the spin Wigner rotation
and so gives the correct (QCD) asymptotic.

\end{abstract}


The relativistic description of bound states was always an
important problem in nuclear physics and particle physics. This
problem became particularly topical in connection with the
development of quark physics, in which the relativistic
properties of the light quarks play a fundamental role.

In the relativistic theory of the description of composite
systems, it is possible to identify two main but very
different approaches.

The first is the method of field theory. Based on the principles
of quantum field theory -- quantum chromodynamics (QCD) -- it is
rightly regarded as the most consistent approach to the solution
of this problem. However, standard perturbative QCD  gives
sufficiently reliable computational prescriptions only for the
description of so--called "hard" processes, which are
characterized by large momentum transfers, and it does not
permit the calculation of characteristics determined by "soft"
processes. Moreover, there are strong indications
\cite{IsL89} that perturbative QCD is not valid for the
description of the currently existing experimental facts in
exclusive processes.  This applies, in particular, to the
description of the elastic form factors of such well--studied
composite systems as the pion, kaon, nucleon and deuteron.  Of
course, in the framework of field theory itself there exist
various approaches to overcoming these difficulties. For
example, there are the well--known approaches associated with
the use of the Bethe--Salpeter equation (see {\it e.g.}
~\cite{JaK90,ItB90,Gro93})
, and
quasipotential approaches (see {\it e.g.}
~\cite{Gro93,BrJ76,LoT63,BlS66,VaD95}).

The second method in the relativistic theory of composite
systems, in the framework of which we shall operate, is based on
the direct realization of the algebra of the Poincar\'e group on
the set of dynamical observables on the Hilbert state space of
the system. This approach is called the theory of direct
interaction, or relativistic Hamiltonian dynamics (RHD)
(for review see ~\cite{KeP91} and references therein).
RHD unifies the potential approach to composite systems and the
condition of Poincar\'e--invariance. It should be noted that the
establishment of the connection between RHD and field theory is
a difficult and as yet unresolved problem. The idea of RHD goes
back to a paper of Dirac ~\cite{Dir49}, in which he considered the
different methods of describing the evolution of classical
relativistic systems -- differing in the evolution parameter:
point form (PF), instant form (IF), and light--front (LF)
dynamics.

There now exists a large number of studies of the use of LF
dynamics
(see, for example,
~\cite{Ter76,BeT76,LeS78,BaK79,KaS92,Coe92,ChC88a,ChC88b,ChC91,
GiK94}
and the references given there).
Some studies also contain investigations using other
form of dynamics.
However, most of quantitative investigations
of specific systems that have so far been made are associated
with LF dynamics. In particular, this is because this form of
dynamics has the smallest number (only 3) of generators that
contain interaction. There are some other advantages that caused
the fact that LF dynamics is widely used.
For example it is the possibility of interpreting the results
with the help of Feynmann diagrams calculated in the
infinite--momentum frame; the antiparticle contributions to
Feynman diagrams are suppressed.
However, the use of the LF dynamics leads to certain
difficulties that are associated with the loss of rotational
invariance ~\cite{Fud90,Kei94}, since the generator of the total angular momentum
contains the interaction. Moreover, the space reflection and
time reversal operators necessarily depend on interactions
~\cite{Lev95}.

Some time ago it was proved that S matrices are equivalent in
the different dynamics forms ~\cite{SoS78}. This fact is interesting but
it does not mean the equivalence of the forms. First, there are
problems which can not be reduced to S matrix, e.g. the
calculation of form factors. Second, one has to keep in mind
that any concrete calculation uses some approximations; the
approximations usually used in different forms of dynamics are
nonequivalent.

Our point of view is the following. One must not be
conservative, one must choose the form adequate to the problem
in question and to the approximation to be done. It seems us
that this is in the spirit of RHD -- the choosing of the
adequate degrees of freedom.

Now we present a relativistic treatment of the problem of soft
electromagnetic structure in the framework of IF of RHD
~\cite{KrT93,BaK94,BaK95,BaK96}. IF of
relativistic dynamics, although not widely used, has some
advantages. The calculations can be performed in a natural
straightforward way without special coordinates. IF is
particularly convenient to discuss the nonrelativistic limit of
relativistic results. This approach is obviously rotational
invariant, so IF is the most suitable for spin problems.

Our approach to electromagnetic structure of two--particle
composite systems has the following advantages.

\begin{itemize}

\item The electromagnetic current matrix element satisfies the
current conservation law
\underline{automatically}.

\item We use relativistic modified impulse approximation
(MIA). It is constructed in relativistic invariant way. This
means that our
\underline{MIA does not} \underline{depend on the choose}
\underline{ of the coordinate
frame}, and this contrasts principally with the
"frame--dependent" impulse approximation usually used in IF
dynamics.

\item Our approach provides with correct and natural
nonrelativistic limit ("the correspondence principle" is
fulfilled).

\item For composite systems (including the spin 1 case) the
approach guarantees the uniqueness of the solution and it does
not use such concepts as "good" and "bad" current components.

\item The approach describes correctly the spin Wigner rotation
and so gives the correct (QCD) asymptotic.

\end{itemize}

It is also worth to notice that our approach is directly linked
with the dispersion approach of quantum field theory
~\cite{TrS69,KuT72,MuT83}  and that it
gives the adequate description of concrete composite systems:
 $\pi-,  K$-- mesons  (quarks systems) and deuteron (nucleons
system).

Let us describe briefly the main steps of investigation
using as an example
quark--antiquark system electromagnetic properties.

Let us consider $\pi$ meson and $K$ meson as
quark ($q$) -- antiquark ($\bar Q$) composite system.
We shall use different quark masses
$M_q$ and  $M_{\bar Q}$ as in $K$ meson.
The results for $\pi$ meson can be obtained if $M_q = M_{\bar Q}$.

The charge form factor for two-quark system can be
obtained from the electromagnetic current matrix element
for composite system
\begin{equation}
<p_c|\,j_\mu\,|{p'}_c>=(p_c+{p'}_c)_\mu\,F_c (Q^2),
\label{pi-par}
\end{equation}
$F_c
(Q^2)$ -- electromagnetic form factor of composite system,
$p_á$ -- 4-mo\-men\-tum of system.

We shall act following the basic assumptions, valid for all forms
of dynamics in RHD \cite{KeP91}.
The RHD
is based on the including of the operator, describing
$q\bar Q$ interaction in the generators of Poincar\'e group while the
commutation relations of Poincar\'e algebra are fulfilled.
One usually
includes $\hat U$ in the mass square operator of the free two particle
system in additive way \cite{KeP91}: $P^2=(p_1+p_2)^2 \>\rightarrow\>{\hat
M}^2_I=P^2+\hat U$. In the case of IF dynamics the Poincar\'e
algebra is conserved if $\hat U$ commutes with the total angular momentum
operator $\hat {\vec J} = ( \hat J_1, \hat J_2, \hat J_3 )$, with the operator
of total 3-momentum $\hat {\vec P}$ and with the operator $\vec \nabla _P$.
The complete set of commuting operators for the two-particle system with
interaction contains now: ${\hat M}^2_I,\>{\hat J}^2,\>\hat J_3,\>\hat
{\vec P}$.
In the case of IF the operators ${\hat J}^2,
\>\hat J_3,\>\hat {\vec P}$ coincide with the
appropriate operators of the two-particle system without interaction and one
can construct the basis in which these three operators are diagonals. While
working in this basis to obtain the wave function one needs to diagonalize
${\hat M}^2_I$.

In RHD the Hilbert space of composite particle states is the tensor
product of single particle Hilbert spaces:
${\cal H}_{q\bar Q} \equiv  {\cal H}_{q} \otimes
{\cal H}_{\bar Q}$
and the state vector in RHD is a superposition of two-particle
states. As a basis in
${\cal H}_{q\bar Q}$ one can choose the following set of vectors:
\begin{eqnarray} |\,\vec p_1\,,m_1;\,\vec p_2\,,m_2\,\!> =
|\,\vec p_1\,,m_1\,\!>\otimes |\, \vec p_1\,,m_2\,>,\nonumber\\ <\,\!\vec
p\,,m\,|\,\vec p\,'\,,
m'\,\!> = 2p_0\,\delta (\vec p - \vec p\,')\,\delta
_{mm'}\>,
\label{RIA}
\end{eqnarray}
Here $\vec p_1 \>,\>\vec p_2$ --- are particle momenta,
$m_1\>,\>m_2$ --- spin projections.

Since we consider the two-quark system as one composite system,
then the natural basis is one with separated center-of-mass
motion:
\begin{equation} |\,\vec P,\>\sqrt
{s},\>J,\>l,\>S,\>m_J\,>\>,
\label{bas-cm}
\end{equation}
with $P_\mu = (p_1 +p_2)_\mu$,
$P^2_\mu = s$, $\sqrt {s}$ ---
the invariant mass of two-particle system , $l$
--- the angular momentum in the center-of-mass frame,
$S$ --- total spin,
$J$ --- total angular momentum, $m_J$ ---
projection of total angular momentum.

The basis (\ref{bas-cm}) is connected with (\ref{RIA})
through the Clebsch -- Gordan decomposition of the Poincar\'e
group.
Now the decomposition of the electromagnetic current matrix
element for the composite system (\ref{pi-par})
in the basis (\ref{bas-cm}) has the form

$$
(p_c + p_c ')_\mu\,F_c (Q^2) = \sum \,\int \,\!\frac {d\vec P}
{N_{C-G}}
\,\frac {d{\vec P}'}
{N'_{C-G}}
\,d\sqrt {s}\,d\sqrt {s'}
<
p_c |\vec P,\sqrt {s},J,l,S,m_J>\cdot $$
\begin{equation}
<\vec P,\sqrt s,J,l,S,m_J \mid j_\mu \mid \vec P',\sqrt{s'}
,J',l',S',{m_J}'>\cdot
\label{pi-cur}
\end{equation}
$$
<\vec P',\sqrt{s'},J',l',S',{m_J}'|{p_c}'>.
$$
Here the sum is over the discrete variables of the basis
(\ref{bas-cm}).\\
$ <\vec P\,,\,\sqrt {s},\,J,\, l,\,S,\,m_J|p_c >$ -
is the composite system wave function
\begin{equation}
<\vec P\,',\,\sqrt
{s'},\,J',\, l',\,S',\,m_J'|\,p_c> =
{N_c}
\delta (\vec P\,' -
\vec p_c)\delta _{JJ'}\delta _{m_Jm_J'}
\delta _{ll'}\delta _{SS'}\,\varphi
^J_{lS}(k)\>.
\label{vf}
\end{equation}
$k = \sqrt{{(s^2-2s(M^2_{\bar s}+M^2_u)+\eta^2)}/{4s}}\>,
\quad \eta = M^2_q-M^2_{\bar Q}. \quad
N_c, N_{C-G}
$
are factors due to normalization. Concrete form of $N_c$ and $N_{C-G}$
will not be used.

Let us discuss the current operator matrix element which
enters the r.h.s. of the equation
(\ref{pi-cur}).

In the case of non-interacting quark system electromagnetic current
matrix element of this system can be parametrized similarly to the standard
case of one-particle matrix element, e.g. for
meson (\ref{pi-par}),
i.e. it is possible to extract the invariant part -- form factor $g_0$:
\begin{eqnarray}
<\vec P,\sqrt s,J,l,S,m_J |\,j^0_\mu\,| \vec P',\sqrt{s'}
,J',l',S',{m_J}'>=\nonumber\\ = A_\mu (s,Q^2,s')\> g_0(s,Q^2,s').
\label{param}
\end{eqnarray}

The vector $A_\mu (s,Q^2,s')$
is defined by the current transformation properties
(by the Lorentz--covariance and the current conservation law):

\begin{equation}
A_\mu =(1/Q^2)[(s-s'+Q^2)P_\mu + (s'-s+Q^2) P\,'_\mu
].  \label{vec-a}
\end{equation}

In our parametrization the
current is conserved by construction:
\begin{equation}
A_\mu(s,Q^2,s') Q^{\mu} = 0.
\label{conserv}
\end{equation}

In the frame of basis (\ref{RIA})
non-interacting current matrix element has the following form:
\begin{eqnarray}
<\vec p_1,m_1;\vec
p_2,m_2
|j^0_\mu|\vec p\,'_1,m'_1;\vec p\,'_2,m'_2>=\nonumber\\
=<\vec p_1,m_1|\vec p\,'_1,m'_1><\vec p_2,m_2|j_\mu|\vec p\,'_2,m'_2>+
  (1\leftrightarrow2).
\label{melRIA}
\end{eqnarray}
This is, as a matter of fact, the relativistic
impulse approximation. The one-particle current in (\ref{melRIA}) is
expressed in terms of one-quark form factors. Clebsh-Gordan
decomposition of the basis (\ref{bas-cm}) into basis (\ref{RIA}) gives the
expression of free form factor $g_0(s,Q^2,s')$ in terms of one-quark form
factors:
$$g_0(s,Q^2,s')=\frac{\sqrt{ss'}}{\sqrt{[s^2-2s(M_{\bar s}^2+M_u^2)+\eta^2]
[s'^2-2s'(M_{\bar s}^2+M_u^2)+\eta^2]}}\cdot $$
\begin{equation}
\label{ff-noint}
\cdot \frac{Q^2(s+s'+Q^2)}{2[\lambda(s,-Q^2,s')]^{3/2}} \cdot
\left (B^u(s,Q^2,s') + B^{\bar s}(s,Q^2,s')\right )\>,
\end{equation}
$$B^{\bar s}(s,Q^2,s') = \left [ f_1^{(\bar s)}(s+s'+Q^2-2\eta)\cos(\omega_1
+ \omega_2)-\right.$$
$$\left. -f_2^{(\bar s)}\frac{M_{\bar s}}{2}\xi (s,Q^2,s')
\sin(\omega_1 +\omega_2 )\right ]\,\theta (s,Q^2,s')\>,$$
$$
\xi (s,Q^2,s') = \sqrt{-\lambda (s,-Q^2,s')M_{\bar s}^2+ss'Q^2-\eta Q^2(s+s'+
Q^2)+Q^2\eta^2}\>,$$
$$\lambda(a,b,c)=a^2+b^2+c^2-2(ab+ac+bc)\>,
$$
$$f_1^{(\bar s)}=\frac{2M_{\bar s}\,G_E^{(\bar s)}(Q^2)}{\sqrt{4M_{\bar
s}^2+Q^2}};\qquad
f_2^{(\bar s)}=-\frac{4\,G_M^{(\bar s)}(Q^2)}{M_{\bar s}\sqrt{4M_{\bar
s}^2+Q^2}}\>,$$
$$\omega_1=\hbox{arctg} \frac{\xi (\,s\,,Q^2\,,s')
}{M_u[(\sqrt{s}+\sqrt{s'})^2+Q^2]
+(\sqrt{s}+\sqrt{s'})(\sqrt{ss'}+\eta)}\>,$$
$$\omega_2=\hbox{arctg} [(\sqrt{s}+\sqrt{s'}+2M_{\bar s})\,
\xi (\,s\,,Q^2\,,s')\cdot$$
$$\{M_{\bar s}(s+s'+Q^2)(\sqrt{s}+\sqrt{s'}+2M_{\bar s})+\sqrt{ss'}
(4M^2_{\bar s}+Q^2)
-\eta [2M_{\bar s}(\sqrt{s}+\sqrt{s'})-Q^2]\}^{-1}]\>,$$
$$\theta (s,Q^2,s') = \vartheta (s'- s_1) - \vartheta (s'- s_2)\>,$$
Here $\vartheta$ is the standard step function, $G_E^{(\bar s)}(Q^2)$
and $G_M^{(\bar s)}(Q^2)$ -- Sachs quark form factors,
$\omega_1$ and $\omega_2$ -- are the Wigner rotation parameters.
$$s_{1,2}=M_{\bar s}^2+M_u^2+\frac{1}{2M_{\bar s}^2}(2M_{\bar
s}^2+Q^2)(s-M_{\bar s}^2-M_u^2) \mp $$
$$\mp\frac{1}{2M_{\bar s}^2}\sqrt{Q^2(4M_{\bar s}^2+Q^2)[s^2-2s(M_{\bar
s}^2+M_u^2)+\eta^2]}\>.$$
Function $B^u(s,Q^2,s')$ can be deduced from $B^{\bar s}(s,Q^2,s')$ by
substitution $M_{\bar s} \leftrightarrow M_u.$

Let us return now to the Eq.(\ref{pi-cur}). The current matrix element
entering the r.h.s. of Eq.(\ref{pi-cur}) must be interaction dependent.
This dependence is known
to be a consequence of the
current conservation law and of the condition of current
relativistic covariance. This means that we can not use
in Eq. (\ref{pi-cur}) the parametrization of non-interacting current
matrix element (\ref{param}) directly and need to include the interaction.
Let us perform the interaction including in (\ref{param}) in minimal
manner:  we shall include the interaction only in the vector function
$A_\mu (s,Q^2,s')$ in Eqs. (\ref{param}), (\ref{vec-a}):
$$ A_\mu (s,Q^2,s')\>\to\> \frac{N_{C-G} N'_{C-G}} {N_cN'_c} A_\mu
^{int}$$

\begin{equation}
A_\mu ^{int}  = A_\mu (s,Q^2,s')\left| _{_{P_\mu \to p_{c \mu},\,\,
P'_\mu \to p'_{c \mu}}} = (p'_c + p_c)_\mu \right.\>,
\label{a-int}
\end{equation}
\begin{equation}
g_0(s,Q^2,s')\>\to\>g(s,Q^2,s') =
g_0(s,Q^2,s')\>.
\label{g-int}
\end{equation}

The current matrix element in Eq.(\ref{pi-cur}) is a
product of a 4-vector and a scalar function
(form factor). This form is quite similar to
the form of electromagnetic current matrix element for two
non-interacting particles (\ref{param}), or to the pion
electromagnetic current matrix element (\ref{pi-par}) and can
differ only by the explicit form of form factors and 4-vectors.
The number of form factors is the same, because all these matrix
elements are taken between the states with $J = l = S = m_J =
0$. Note, that all the normalization constants, which are not
invariant, are included in the covariant part.

Let us rewrite the equation (\ref{pi-cur}) using meson wave
function (\ref{vf}) and current matrix element explicitly:
\begin{equation}
(p_c + p_c')_\mu\,F_c(Q^2) =
\int\,d\sqrt{s}\,d\sqrt{s'}\,\varphi(k)\,
A_\mu^{int}(s,Q^2,s')\,g(s,Q^2,s')\,\varphi(k')\>.
\label{re}
\end{equation}
Here we use for simplicity the notation:
$ \varphi ^J_{lS}(k)\> \to \> \varphi (k). $

This means that the two 4-vectors are equal and this equality is
to be valid for any choice of wave functions $\varphi(s)$ of
the two-particle system internal motion.  If the wave function
is varied then the scalar part of the l.h.s. (the form factor
$F_c(Q^2)$) is changed, while the covariant part (the vector
$(p_c + p_c')_\mu$) remains unchanged, because the vector $(p_c
+ p_c')_\mu$ describes the system as a whole and does not depend
on the interaction inside the system. So, when the wave function
is varied the l.h.s. remains to be collinear to the vector
$(p_c + p_c')_\mu$. In general case the 4-vector in the r.h.s.
changes the direction. The equality is valid for an arbitrary
choice of wave function only if the vector
$A_\mu^{int}$ is collinear to the vector $(p_c + p_c')_\mu$
in any coordinate system, so that the proportionality
factor can be included in the invariant form factor
$g(s,Q^2,s')$. Thus the form (\ref{a-int}) for $A_\mu^{int}$ is
unique and the most general.

The choice (\ref{g-int}) for the form factor $g(s,Q^2,s')$ is
not quite general, of course. One can use different physical
approximations to evaluate this quantity. The use of
$g_0(s,Q^2,s')$ (\ref{ff-noint}) instead of $g(s,Q^2,s')$ means
relativistic impulse approximation as formulated mathematically
in terms of form factors (modified impulse approximation).

The function $A_\mu ^{int}$
contains the interaction through the impulses $p'_{c \mu}$ and $p_{c
\mu}$.  Using (\ref{pi-cur}), (\ref{param}), (\ref{a-int})
and (\ref{g-int})
we obtain now the following expression for the form factor:
\begin{equation}
F_c (Q^2)=\int d\sqrt s\ d\sqrt{s'}\ \varphi (k)\,g_0(s,Q^2,s')\,
\varphi (k').
\label{ff}
\end{equation}

For $\varphi (k)$ one can use  any phenomenological
wave function, normalized using the relativistic density of
states:
$\varphi(k) =\sqrt{\sqrt{s}(1 - \eta^2/s^2)}\,u(k)\,k,$
$u(k)$ - is nonrelativistic phenomenological wave function.

Let us emphasize, that
the r.h.s. of Eq.(\ref{pi-cur}) with (\ref{a-int}) inserted
satisfies the current conservation law: it is orthogonal to the
vector ${Q_\mu = (p'_c - p_c)_\mu}$. This latter fact is rather
noticeable because generally the construction of the
conserved current operator for composite systems presents
a rather complicated problem which is not solved yet
\cite{Lev95}.
Thus, the Eq.(\ref{ff}) takes into account the relativistic
covariance and the current conservation law. This is right for
any choice of the function $g(s,\,Q^2,\,s')$,
including the expressions (\ref{ff-noint}), (\ref{g-int}) which
we use here.

We have obtained the expression (\ref{ff}) for the form factors
in the frame of the principally relativistic approach: instant
form of RHD. The form factors are expressed in terms of
relativistic function $g(s,\,Q^2,\,s')$ and nonrelativistic
wave functions $u(k)$. The behavior of form factor depends
essentially on the model type of wave function. The RHD instant
form enables one to obtain easily the nonrelativistic limit of
Eq.(\ref{ff}). The relativistic effects are important: the
difference between relativistic and nonrelativistic form
factors for one and the same wave function is very large.

To conclude, we present an approach -- the modified IF of RHD --
which uses relativistic parametrization and so picks out the
relativistic invariant quantities on each stage of the
calculation. The approach describes well the data on
electromagnetic form factors of
$\pi-, K-$ mesons and the deuteron
(see the authors' poster at FBXV).

V.E.T.\ is indebted to Ben Bakker for the possibility to attend
the Workshop on Relativistic Approaches to Few-Body Systems and 
to present these results. This work is
supported in part by the Russian Foundation for Basic Research
(grant no.~96-02-17288).


\newpage

\end{document}